\documentclass[conference]{IEEEtran}
\IEEEoverridecommandlockouts
\usepackage{cite}
\usepackage{amsmath,amssymb,amsfonts}
\usepackage{algorithm}
\usepackage{algpseudocode}
\usepackage{graphicx}
\usepackage{textcomp}
\usepackage{xcolor}
\usepackage{booktabs}
\usepackage{multirow}
\usepackage{url}
\usepackage{tikz}
\usetikzlibrary{arrows.meta, positioning, calc, fit, backgrounds}

\def\BibTeX{{\rm B\kern-.05em{\sc i\kern-.025em b}\kern-.08em
    T\kern-.1667em\lower.7ex\hbox{E}\kern-.125emX}}

\begin{document}

\title{Probing the Prefill: Detecting Code Vulnerabilities via Latent Activations}

\author{
\IEEEauthorblockN{Alizishaan Khatri}
\IEEEauthorblockA{\textit{Wrynx Inc.} \\
research@wrynx.com}
}

\maketitle

\begin{abstract}
LLM-based code generation is now embedded in mission-critical
pipelines, but defenses against vulnerable output remain post-hoc --
static analyzers, fine-tuned classifiers, or an LLM judge that screen
\emph{completed} code, ignoring the generating model's own internal
state. We test a narrower, directly measurable question: when an LLM
reads a piece of C/C++ code as context, do its hidden activations
already carry a signal, recoverable easily, about that code's vulnerability status? We extract last-prefill-token activations
from four LLMs (Granite-4.1-8B, Qwen3.5-9B, Qwen3.6-27B, Gemma-4-12B)
across three model families and train MLP probes on these activations,
evaluated on four function-level C/C++ benchmarks (Devign, Big-Vul,
Draper VDISC, PrimeVul) using each dataset's existing, human/CVE-labeled
functions as the probed context -- not model-generated completions,
which we have not yet tested (\S\ref{sec:methodology}). So this is
direct evidence for activation-level informativeness and only indirect
motivation for the stronger claim that a model could flag its own
forthcoming output as vulnerable. Our probes achieve 41.7\% average F1
using 13.4--16.0M-parameter probes -- under 0.2\% of base-model size.
On Devign, the best probe (Qwen3.5-9B, 68.8\% F1) matches the
published fine-tuned-classifier SOTA (67.9\%) despite reading only a
frozen, general-purpose LLM's activations; on the harder, more
imbalanced benchmarks (Big-Vul, Draper VDISC, PrimeVul) probes trail
SOTA substantially, tracking those datasets' known label-imbalance and
quality problems rather than an absence of signal. This is early
evidence that a coding LLM's own representation of arbitrary code is
informative about that code's vulnerability status, motivating further
work toward lightweight, model-native vulnerability screening.
\end{abstract}


\section{Introduction}

LLM-based coding assistants and autonomous coding agents are now used
to generate production code, including in defense and safety-critical
pipelines \cite{app14031046} that demand strong assurance against
vulnerable output \cite{pmlr-v284-sevenhuijsen25a}, across
applications such as autonomous vehicles \cite{nouri2025simulation}
and self-organizing robot swarms \cite{zhu2025online}. As with
harmful-content generation, the dominant defenses against vulnerable
code generation operate at the interface -- static analyzers, learned
classifiers, or a second LLM run over the \emph{completed} code -- and
share the structural weaknesses common to any post-hoc, black-box
screen (\S\ref{sec:relatedwork}). A parallel line of interpretability
work argues this asymmetry is unnecessary: LLMs already encode
substantial task-relevant semantics, including whether a prompt is
harmful, as linearly separable directions in their hidden states
\cite{pmlr-v235-zheng24n, jiao-etal-2026-llm, sternfeld2026minimal,
khatri2026safety, khatri2026all} (\S\ref{sec:relatedwork}).

We ask: \emph{When an LLM reads a piece of code as context, does its
internal state already encode a recoverable signal about that code's
vulnerability status, consistently enough across model families to
motivate testing it as a generation-time gate?} We test this on
existing, human/CVE-labeled corpus functions, not model-generated
completions (\S\ref{sec:methodology}) -- a necessary but not sufficient
step toward a model flagging its own forthcoming vulnerable output,
which we do not test here. This is a preliminary, workshop-scale study;
\S\ref{sec:limitations} enumerates the controls (an architecture
ablation, a non-LLM baseline, an adversarial-robustness check) still
needed to fully support the claims below.

\textbf{Contributions.}
\begin{itemize}
\item We extend activation probing for LLM safety (\cite{khatri2026safety, saglam-etal-2025-large} and related probing work \cite{ibanezlissen2025lpass,sternfeld2026minimal}) to \emph{code vulnerability detection}, rather than prompt-harm classification, where code appeared only incidentally as a minority of Aegis-dataset samples \cite{ghosh-etal-2025-aegis2} (\S\ref{sec:relatedwork}).
\item We extract last-prefill-token activations from four LLMs
(Granite-4.1-8B, Qwen3.5-9B, Qwen3.6-27B, Gemma-4-12B) across three
model families and train MLP probes to predict the vulnerability label
of the code in context, evaluated on four function-level C/C++
vulnerability benchmarks -- Devign, Big-Vul, Draper VDISC, and PrimeVul
(\S\ref{sec:methodology}).
\item We report precision, recall, and F1 for every (model, dataset)
pair (\S\ref{sec:results}).
\item For all four probes on Big-Vul, we break down classification
accuracy by CWE type and find it uniformly high and tightly clustered
across the five most frequent types, with CWE-119 the single hardest
category for every model (\S\ref{sec:results}).
\item We discuss the latency/compute case for probe-based
generation-time gating over post-hoc baselines, and known
generalization failure modes of probing-based detectors
\cite{wang2025falsesense,fomin2026benchmarkslie} (\S\ref{sec:limitations}).
\end{itemize}

\section{Related Work}
\label{sec:relatedwork}

\subsection{Post-Hoc Vulnerability Detection}
Most production screening for LLM-generated code runs static/dynamic
analyzers over the output or fine-tunes a classifier, often another
LLM, on labeled code \cite{steenhoek2023languagemodels}. Such
detectors are \emph{post-hoc}, \emph{black-box} to the generating
model's own computation, and evadable by surface transformations that
preserve the underlying vulnerability \cite{li2025obfuscation}.
Steenhoek et al.\ \cite{steenhoek2023languagemodels} show via
attention analysis that detection models already encode bug-relevant
semantics, though not as a linear probe. \textbf{We probe that
internal state directly via the model's own forward pass, rather than
analyzing emitted code after the fact.}

\subsection{Probing LLM Internals for Vulnerability Detection}
LPASS \cite{ibanezlissen2025lpass} places linear probes after every
layer of a compressed LLM to guide pruning, for compression, not
screening. Sternfeld et al.\ \cite{sternfeld2026minimal}, our closest
prior work, train logistic-regression/MLP probes on \emph{prompt-end}
hidden states to predict whether a yet-to-be-generated completion will
be secure, showing minimal prompt perturbations can flip it.
\textbf{We differ}: deeper 6-layer MLP probes; activation over the
candidate code itself from four dedicated corpora, not a prompt paired
with a generated completion; and an emphasis on cross-model (4 LLMs, 3
families) detection rather than perturbation sensitivity.

\subsection{Latent-State Probing Beyond Code}
Latent-state probing has also targeted malicious prompts, jailbreaks,
and malware. Chia et al.\ \cite{chia2025probinglatent} separate safe
from jailbroken states in latent subspaces; Fomin
\cite{fomin2026benchmarkslie} probes three LLMs under a
leave-one-dataset-out protocol; Wang et al.\ \cite{wang2025falsesense}
show such probes generalize poorly out-of-distribution
(\S\ref{sec:relatedwork-malware}); Ajayi et al.\ \cite{ajayi2025vae}
apply VAE latent spaces to malware detection outside the LLM setting.

\section{Problem Formulation}
\label{sec:problem}

Let $\mathcal{M}$ be an LLM used for code generation and
$\mathcal{V}: \mathcal{O} \to \{0,1\}$ a ground-truth vulnerability
oracle over completions $o = \mathcal{M}(x)$ for coding prompt $x$,
such that $\mathcal{V}(o) = 1$ iff $o$ contains an instance of one or more target vulnerability classes (e.g., a CWE, as determined by dataset ground truth. We define the last-prefill-token activation
$h(x) = \mathcal{M}_{\text{hidden}}(x)[-1] \in \mathbb{R}^{d}$ as the
final-layer hidden state at the last prompt token, immediately before
generation begins. We seek a probe $f_\theta: \mathbb{R}^{d} \to
\{0,1\}$, trained on frozen activations $h(x)$, that predicts
$\mathcal{V}(\mathcal{M}(x))$ \emph{before} $\mathcal{M}(x)$ is
generated -- unlike post-hoc detectors, which operate on $o$ directly,
and unlike the harmful-prompt setting of \cite{khatri2026safety},
where the probe predicts a property of $x$ itself. This is the target
formulation; none of our four evaluation datasets pairs a prompt $x$
with a model-generated completion $o=\mathcal{M}(x)$, so
\S\ref{sec:methodology} operationalizes it by substituting an existing,
dataset-labeled function $c$ for $x$ and reading $\mathcal{V}(c)$
directly off the dataset -- a proxy for, not a direct test of, the
$\mathcal{V}(\mathcal{M}(x))$ formulation above. Fig.~\ref{fig:pipeline}
and \S\ref{sec:trainingprocedure} detail the resulting extraction and
training pipeline; the practical objective is
a probe cheap enough to run inline with generation ($T_{\text{probe}}
\ll T_{\text{llm}}$) that generalizes across datasets and model
families rather than overfitting to one model's idiosyncratic
representation of vulnerability.

\section{Methodology}
\label{sec:methodology}

\subsection{Overview}
Our pipeline mirrors the two-stage design of
\cite{khatri2026safety}: (1) an \emph{extraction} stage that runs each
coding prompt through a frozen LLM and stores the last-prefill-token
activation, and (2) a \emph{training} stage that fits an MLP probe on
those frozen activations to predict vulnerability of the resulting
completion.

\begin{figure}[t]
\centering
\resizebox{\linewidth}{!}{%
\begin{tikzpicture}[
  box/.style={draw, rounded corners=2pt, align=center, minimum width=2.9cm,
    minimum height=8mm, font=\scriptsize, inner sep=2.5pt, fill=white, thick},
  frozenbox/.style={box, dashed, draw=blue!55!black, fill=blue!4},
  trainbox/.style={box, draw=orange!70!black, fill=orange!7},
  arr/.style={-{Latex[length=1.8mm]}, thick},
  lbl/.style={font=\scriptsize, midway, fill=white, inner sep=1pt}
]

\node[box]       (func)  at (-2.1, 0)     {Candidate function $c$\\(from dataset)};
\node[frozenbox] (llm)   at (-2.1, -1.05) {Frozen target LLM $\mathcal{M}$};
\node[box]       (act)   at (-2.1, -2.10) {Activation $h(c)\in\mathbb{R}^d$\\(last-prefill token)};
\node[box]       (cache) at (-2.1, -3.15) {Cached pairs $(h_i, y_i)$};

\draw[arr] (func) -- node[lbl]{forward pass} (llm);
\draw[arr] (llm) -- (act);
\draw[arr] (act) -- node[lbl]{paired w/ label $y$} (cache);

\node[trainbox] (probe) at (2.1, 0)     {MLP probe $f_\theta$\\(trainable)};
\node[box]      (pred)  at (2.1, -1.05) {Prediction $\hat{y}\in\{0,1\}$};
\node[box]      (loss)  at (2.1, -2.10) {Class-weighted BCE loss};

\draw[arr] (cache.east) -- (0, -3.15) -- (0, 0) -- (probe.west);
\draw[arr] (probe) -- (pred);
\draw[arr] (pred) -- (loss);
\draw[arr, dashed] (loss.east) to[out=-15, in=15, looseness=2.6]
  node[right, font=\scriptsize, align=left]{backprop\\($\theta$ only)} (probe.east);

\begin{scope}[on background layer]
\node[draw=blue!45, rounded corners, fill=blue!2, thick,
  fit=(func)(llm)(act)(cache), inner sep=3.5mm,
  label={[font=\bfseries\scriptsize, blue!55!black]above:Stage 1: Extraction}] {};
\node[draw=orange!55, rounded corners, fill=orange!2, thick,
  fit=(probe)(pred)(loss), inner sep=3.5mm,
  label={[font=\bfseries\scriptsize, orange!65!black]above:Stage 2: Training}] {};
\end{scope}

\end{tikzpicture}%
}
\caption{Two-stage probing pipeline (cf.\ Fig.~2 of \cite{khatri2026safety}).
\textbf{Stage 1 (Extraction):} each function $c$ passes through the frozen
LLM $\mathcal{M}$ (dashed border denotes frozen weights) and its
last-prefill-token activation $h(c)\in\mathbb{R}^d$ is cached with label $y$.
\textbf{Stage 2 (Training):} an MLP probe $f_\theta$ is trained on the
cached $(h_i, y_i)$ pairs via a class-weighted binary cross-entropy
loss (\S\ref{sec:trainingprocedure}); only $\theta$ is updated --
$\mathcal{M}$ stays frozen.}
\label{fig:pipeline}
\end{figure}
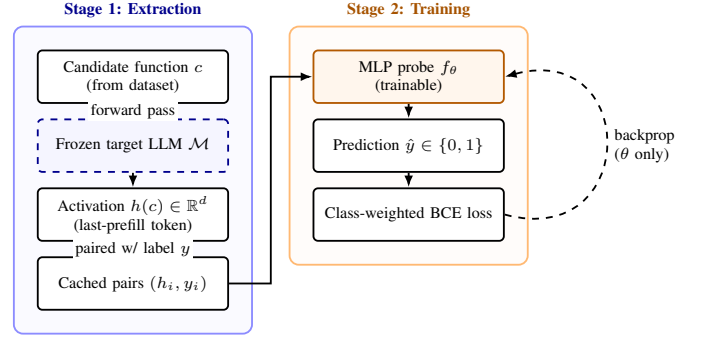

\subsection{Models}
We evaluate probes trained on activations from four LLMs spanning
three publishers, to test whether vulnerability-relevant
representations are a general property of coding-capable LLMs rather
than an artifact of a single model, and, via the two Qwen sizes, to
probe within-family generalization across a scale/generation jump:
\textbf{Granite-4.1-8B} (IBM, dense decoder-only) \cite{ibm2026granite41};
\textbf{Qwen3.5-9B} (Alibaba, hybrid gated-DeltaNet/gated-attention)
\cite{qwen2026qwen35}; \textbf{Qwen3.6-27B} (Alibaba, same family as
Qwen3.5-9B, later generation) \cite{qwen2026qwen36}; and
\textbf{Gemma-4-12B} (Google DeepMind, encoder-free decoder-only)
\cite{google2026gemma4}.

Table~\ref{tab:modelspecs} lists the hidden dimension $d$ and layer
count for each model, which fixes the input width of that model's
probe (Table~\ref{tab:architecture}).

\begin{table}[t]
\centering
\footnotesize
\caption{Target Model Specifications}
\label{tab:modelspecs}
\begin{tabular}{@{}lccc@{}}
\toprule
Model & Params & $d$ (hidden dim) & Layers \\
\midrule
Granite-4.1-8B & 8B & 4096 & 40 \\
Qwen3.5-9B & 9B & 4096 & 32 \\
Qwen3.6-27B & 27B & 5120 & 64 \\
Gemma-4-12B & 12B & 3840 & 48 \\
\bottomrule
\end{tabular}
\end{table}

For each model, we extract the final hidden layer's last-prefill-token
activation using last-token pooling, following
\cite{khatri2026safety}. We probe only the final layer's last-token
representation (a single $d$-dimensional vector per example), not a
layerwise sweep as in \cite{ibanezlissen2025lpass}; \S\ref{sec:limitations}
notes this as a direction their layer-pruning results motivate for
future work.

\subsection{Datasets}
We evaluate on four widely used function-level vulnerability detection
datasets, all C/C++ and all binary-labeled (vulnerable / benign) at
function granularity, chosen to span a range of labeling
methodologies (manual security-expert review, CVE/patch mining, and
static-analyzer-derived labels) and known label-quality regimes,
since label noise is itself an active concern in this literature
\cite{ding2024primevul}:

\begin{table}[t]
\centering
\footnotesize
\caption{Dataset Summary and Reported Binary-Classification SOTA}
\label{tab:datasets}
\begin{tabular}{@{}lp{0.75cm}p{1.05cm}cp{1.55cm}@{}}
\toprule
Dataset & Size & Vuln.\ \% & SOTA F1 & Labeling \\
\midrule
Devign \cite{zhou2019devign} & 27K & 46\% & 67.9\% & manual review \\
Big-Vul \cite{fan2020bigvul} & 217K & 5.8\% & 96.5\%* & CVE/patch mining \\
Draper VDISC \cite{russell2018draper} & 1.27M & $\sim$9.6\% & 60.9\% & static-analyzer \\
PrimeVul \cite{ding2024primevul} & 236K & $\sim$3\% & 24.5\%* & auto-labeling \\
\bottomrule
\end{tabular}
\end{table}

All four are C/C++, function-level, binary-labeled (vulnerable/benign).
\textbf{Devign} \cite{zhou2019devign} (HF \texttt{DetectVul/devign},
27.4K rows: 21.9K/2.7K/2.7K train/val/test) draws from the Linux
kernel, QEMU, Wireshark, and FFmpeg, manually labeled; relatively
balanced ($\sim$46\%, smaller than other Devign mirrors in
circulation); SOTA is 67.9\% F1 (SNOPY \cite{risse2025topscore}).
\textbf{Big-Vul} \cite{fan2020bigvul} (HF \texttt{bstee615/bigvul})
links 309 GitHub projects to 2002--2019 CVEs via patch commits;
imbalanced, known label noise \cite{ding2024primevul}; reported SOTA
is 96.5\% F1 (DeepDFA \cite{risse2025topscore}), but a bag-of-words
classifier with no code structure reaches 86\% F1 in the same study --
treat this SOTA with caution. \textbf{Draper VDISC} \cite{russell2018draper}
(HF \texttt{claudios/Draper}) labels via a static-analysis ensemble
(Clang SA, Cppcheck, Flawfinder); 80/10/10 split; SOTA is 60.9\% F1
(CodeT5 \cite{norton2026vulnbench}). \textbf{PrimeVul}
\cite{ding2024primevul} (HF \texttt{colin/PrimeVul}) spans 140+ CWE
types, purpose-built to fix label noise/contamination in Devign and
Big-Vul via improved auto-labeling, deduplication, and chronological
splits; best reported F1 when actually fine-tuned and evaluated on
PrimeVul is 24.5\% (UniXcoder with class-aware contrastive learning
\cite{ding2024primevul}), still far below the other three benchmarks.
The same study reports a starker number: a 7B model fine-tuned on
\emph{Big-Vul} scores 68.3\% F1 there but only 3.1\% when that
checkpoint is evaluated on PrimeVul's test set -- a cross-dataset
transfer collapse, not PrimeVul's achievable SOTA, but the starkest
sign of how gameable the other three benchmarks are, and why we
evaluate across all four rather than Devign/Big-Vul alone.

\textbf{Operationalization.} As \S\ref{sec:problem} flags, we
substitute each dataset function $c$ for $x$ and pool $h(c)$ (the same
forward pass, run as input context) rather than pairing a prompt with
a generated completion, avoiding the post-hoc completion labeling of
\cite{sternfeld2026minimal} and instead treating the LLM as a fixed
feature extractor, as in LPASS \cite{ibanezlissen2025lpass}. The
forward pass is identical whether $c$ is a corpus function (evaluated
here) or a completion the model just emitted, but we have not yet run
the latter (\S\ref{sec:limitations}).

\subsection{Probe Architecture}
Following \cite{khatri2026safety}, we use an MLP with progressively
decreasing hidden dimensions, GELU activations, and dropout
(Table~\ref{tab:architecture}). We depart from \cite{khatri2026safety}
in the output layer: rather than its softmax over two classes, ours
ends in a single raw logit trained with a class-weighted binary
cross-entropy loss (\S\ref{sec:trainingprocedure}), matching our
binary vulnerable/benign task rather than the multi-class-style head
their prompt-harm setting used. Because our four models do not share a
hidden dimension $d$ (Table~\ref{tab:modelspecs}), we instantiate one
probe per model with its own input width $d$ but an otherwise
identical architecture, so cross-model comparisons isolate the
representation rather than probe capacity. This 6-layer architecture
is inherited from \cite{khatri2026safety}'s harmful-prompt task, not
independently justified for code, and widths/dropout ($p=0.1$) were
fixed a priori rather than swept; we have not tested a linear
(logistic-regression) probe on the same activations to see whether the
nonlinear capacity is earned (\S\ref{sec:limitations}) -- so
``recoverable'' claims about our results should not be read as
evidence of linear separability specifically.

\begin{table}[t]
\centering
\footnotesize
\caption{MLP Probe Architecture (Shared Across Models; Only $d$ Varies)}
\label{tab:architecture}
\begin{tabular}{@{}cccc@{}}
\toprule
Layer & In & Out & Activation \\
\midrule
1 & $d$ (Table~\ref{tab:modelspecs}) & 2048 & GELU + Dropout($p{=}0.1$) \\
2 & 2048 & 2048 & GELU + Dropout($p{=}0.1$) \\
3 & 2048 & 512 & GELU + Dropout($p{=}0.1$) \\
4 & 512 & 512 & GELU + Dropout($p{=}0.1$) \\
5 & 512 & 64 & GELU + Dropout($p{=}0.1$) \\
6 & 64 & 1 & (raw logit; sigmoid applied by the loss) \\
\bottomrule
\end{tabular}
\end{table}

\subsection{Training Procedure}
\label{sec:trainingprocedure}
Probe training is adapted from \cite{khatri2026safety} to predict
completion-level vulnerability rather than prompt-level harm: cache
each $(h_i, y_i)$ pair (Fig.~\ref{fig:pipeline}), then optimize
$f_\theta$ by class-weighted binary cross-entropy, checkpointing and
threshold-sweeping every epoch as detailed below. We train with AdamW
\cite{loshchilov2019decoupled} ($\eta = 2.5\times10^{-4}$, weight decay
$10^{-2}$, $\beta = (0.9, 0.9999)$, $\epsilon = 10^{-8}$), batch size
1024, for up to 50 epochs, with a ReduceLROnPlateau schedule that
halves $\eta$ after 2 epochs without a $\geq 10^{-2}$ improvement in
validation loss. All four datasets are skewed toward the benign class
(Table~\ref{tab:datasets}), most severely PrimeVul ($\sim$3\% vulnerable);
we address this with a class-weighted binary cross-entropy loss whose
positive-class weight is $(n_{\text{neg}}/n_{\text{pos}})^{p}$ on the
training split ($p=1$ full inverse-frequency reweighting by default,
with $p=0.5$ available as a gentler compromise), optionally combined
with minority-class oversampling to a configurable target ratio
(used for PrimeVul in our runs). Because reweighting shifts the
probe's output distribution, the fixed 0.5 sigmoid cutoff is no longer
guaranteed to be the best decision boundary: after every epoch we
sweep the threshold on validation to maximize F1 (configurable to
accuracy/precision/recall) and apply the resulting fixed threshold to
the test split, reporting the AUC (threshold-independent) alongside.
Checkpoints are saved every epoch, and the best checkpoint is selected
by validation loss by default (configurable to the validation F1 at
the tuned threshold instead); \S\ref{sec:results} reports precision,
recall, and F1 at that checkpoint's tuned threshold where the
threshold sweep for that specific checkpoint is available, and at the
untuned 0.5 cutoff otherwise (marked in Table~\ref{tab:results}).

\subsection{Evaluation Metrics}
For each (model, dataset) pair we report precision, recall, F1, and
AUC on a held-out test split, following \cite{khatri2026safety}. We
additionally break down accuracy per CWE type for the two datasets
whose schema exposes CWE labels -- Draper VDISC (per-CWE boolean
columns) and Big-Vul (regex extraction from a free-text field) -- at a
checkpoint's tuned threshold; Devign/PrimeVul lack a schema-specific
extractor. We do not yet report a cross-model/cross-dataset
generalization matrix, which \S\ref{sec:relatedwork-malware} and
\cite{wang2025falsesense} motivate (\S\ref{sec:limitations}).

\subsection{Experimental Setup}
\label{sec:setup}
\textbf{Hardware:} 2$\times$ NVIDIA L4 and 2$\times$ NVIDIA A100 GPUs,
2 CPUs, and 1TB of storage (local SSD plus cloud). GPUs were used
exclusively for extracting activations from the target LLMs; probe
training and visualization (\S\ref{sec:trainingprocedure},
Table~\ref{tab:cwe-accuracy}) ran on CPU only.
\textbf{Splits:} We use each dataset's native HuggingFace
train/validation/test split as-is (Table~\ref{tab:datasets}); we do not
apply custom stratification by label or CWE class on top of the
dataset-provided split.

\section{Results}
\label{sec:results}

Table~\ref{tab:results} reports probe performance across all 16
(model, dataset) pairs, averaging 41.7\% F1 overall. Difficulty splits
by \emph{dataset}, not model: Devign is easiest (62.1--68.8\% F1) and
roughly matches its published SOTA (67.9\%, Table~\ref{tab:datasets})
for three of four models; Big-Vul is next (44.6--59.5\%) but well
below its likely-inflated 96.5\% SOTA (\S\ref{sec:methodology}); Draper
VDISC and PrimeVul are both hard and strikingly
\emph{model-invariant} -- 30.5--30.9\% and 17.6--19.5\% F1
respectively, despite base models differing in size by over
3$\times$. This tightness is consistent with a hypothesis that dataset
label quality and class imbalance, not which model's activations the
probe reads, set the ceiling on these two benchmarks -- also
consistent with PrimeVul's own SOTA
(24.5\%,~\cite{ding2024primevul}) needing purpose-built imbalance
handling, not a different architecture -- but we have not run the
non-LLM baseline (e.g.\ TF-IDF/AST) that would confirm this rather
than the probes simply underperforming an achievable ceiling
(\S\ref{sec:limitations}). No model is uniformly best across all four
datasets, though several cells (marked $^\dagger$) use an untuned
threshold, so part of this ranking may reflect threshold availability
rather than a genuine model difference.

\begin{table*}[t]
\centering
\small
\caption{Probe Performance Across Models and Datasets (Precision~/~Recall~/~F1, \%)}
\label{tab:results}
\begin{tabular}{@{}lccc ccc ccc ccc@{}}
\toprule
& \multicolumn{3}{c}{Devign} & \multicolumn{3}{c}{Big-Vul} & \multicolumn{3}{c}{Draper VDISC} & \multicolumn{3}{c}{PrimeVul} \\
\cmidrule(lr){2-4} \cmidrule(lr){5-7} \cmidrule(lr){8-10} \cmidrule(lr){11-13}
Model & P & R & F1 & P & R & F1 & P & R & F1 & P & R & F1 \\
\midrule
Granite-4.1-8B & 58.15 & 75.63 & 65.74 & 43.26 & 46.00 & 44.59 & 24.15 & 41.48 & 30.53$^\dagger$ & 27.48 & 15.12 & 19.51$^\dagger$ \\
Qwen3.5-9B & 56.30 & 88.35 & 68.77 & 44.69 & 88.99 & 59.50$^\dagger$ & 22.05 & 50.27 & 30.65$^\dagger$ & 15.77 & 19.85 & 17.58 \\
Qwen3.6-27B & 58.38 & 70.28 & 63.78 & 43.73 & 52.73 & 47.81 & 23.33 & 45.78 & 30.91$^\dagger$ & 23.80 & 15.30 & 18.63 \\
Gemma-4-12B & 58.10 & 66.76 & 62.13 & 42.50 & 88.89 & 57.50$^\dagger$ & 22.24 & 48.47 & 30.49$^\dagger$ & 23.35 & 15.48 & 18.62$^\dagger$ \\
\bottomrule
\end{tabular}
\\[2pt]
\raggedright\footnotesize $^\dagger$Untuned 0.5-threshold result
(validation-F1 sweep unavailable for this checkpoint); P/R are not
directly comparable to undaggered cells in the same column
(\S\ref{sec:trainingprocedure}).
\end{table*}

\subsection{Per-Vulnerability-Class Breakdown}
For Big-Vul, we group each test-split row by CWE type (not mutually
exclusive) and report accuracy at the tuned threshold within the five
most frequent types (of 85 total; the rest are too sparse), plus a
residual ``Benign / No CWE'' bucket; Draper VDISC is supported by the
same tooling but not yet run, and Devign/PrimeVul lack the needed
schema. Table~\ref{tab:cwe-accuracy}: accuracy is uniformly high and
tightly clustered, 97.5--99.0\% across all five types and within about
1 point of each model's Benign/No-CWE accuracy. \textbf{CWE-119
(buffer-bounds violations, $n=4478$) is the single hardest category
for all four models}, and CWE-416/CWE-399 are consistently easiest. We
read this uniformity cautiously: Big-Vul skews heavily benign
(Table~\ref{tab:datasets}), all four checkpoints use a high tuned
threshold (0.919--0.945), and its reported SOTA is itself inflated by
a bag-of-words baseline reaching 86\% F1 \cite{risse2025topscore} --
near-ceiling accuracy here is consistent both with CWE-independent
vulnerability signal and with a general vulnerability-correlated
shortcut; accuracy alone cannot distinguish the two.

\begin{table}[t]
\centering
\footnotesize
\caption{Per-CWE-Type Classification Accuracy (\%) on Big-Vul, Top-5
CWE Types Plus the Benign/No-CWE Bucket}
\label{tab:cwe-accuracy}
\begin{tabular}{@{}lcccc@{}}
\toprule
CWE Type ($n$) & Granite & Qwen3.5 & Qwen3.6 & Gemma \\
\midrule
Benign / No CWE (6424) & 98.8 & 98.6 & 98.9 & 98.3 \\
CWE-119 (4478) & 97.9 & 97.5 & 97.7 & 97.7 \\
CWE-20 (3388)  & 98.1 & 97.8 & 98.0 & 97.8 \\
CWE-399 (2889) & 98.8 & 98.6 & 98.7 & 98.6 \\
CWE-264 (2287) & 98.1 & 98.3 & 98.3 & 98.6 \\
CWE-416 (1559) & 99.0 & 98.8 & 98.8 & 98.3 \\
\bottomrule
\end{tabular}
\\[2pt]
\raggedright\footnotesize Tuned threshold per model: Granite-4.1-8B
$\tau{=}0.922$; Qwen3.5-9B $\tau{=}0.926$; Qwen3.6-27B $\tau{=}0.945$;
Gemma-4-12B $\tau{=}0.919$.
\end{table}

\section{Discussion}
\label{sec:discussion}

\subsection{Why Probe at Generation Time: The Resource-Savings Case}
\label{sec:resource-savings}
The case for probing over external screening is fundamentally a
resource argument: \cite{khatri2026safety}'s 12.6M-parameter probe is
0.16\% of the 8B-parameter prompt-harm model it reads from, yet
matches 7B-parameter guard models (82.7\% F1 vs.\ WildGuard's 84.4\%
on BeaverTails) and trails GPT-4-scale judges by only a few points.
Two savings compound:

\begin{itemize}
\item \textbf{Parameter cost.} A probe adds a negligible fraction of
the base model's parameters versus deploying a second 7B--1.8T model
as an external judge. Our probes range 13.41--16.03M parameters
(Table~\ref{tab:architecture}), i.e.\ 0.17\%--0.06\% of the respective
base model -- the same order of magnitude as \cite{khatri2026safety}'s
0.16\%, and \emph{shrinking} as the base model scales up, since
absolute probe size barely grows with $d$.
\item \textbf{Latency and infrastructure.} External screening is
\emph{additive} (prompt filter, generation, response filter run
sequentially, each potentially a separate GPU service with network
round-trips); a probe reading the primary model's own activations is
\emph{concurrent} and needs no separate infrastructure, bounding total
latency by the slower of $T_{\text{llm}}$ and $T_{\text{probe}}$.
Since probe parameters are a small fraction of primary-model
parameters, $T_{\text{probe}} \ll T_{\text{llm}}$ in practice --
\cite{khatri2026safety} measured under 1\,ms against 50--500\,ms
generation times, effectively free -- and, reading prefill-time
activations, can in principle flag risk before a vulnerable completion
is fully emitted, unlike a response filter that must wait for it.
\end{itemize}

\textbf{Relevance to tactical and air-gapped deployment.} Beyond raw
resource savings, a same-process probe needs no round-trip to an
external judge, and hence no network path for generated code --
itself potentially sensitive or classified
\cite{app14031046,pmlr-v284-sevenhuijsen25a} -- to leave the inference
environment. This suits on-device or air-gapped pipelines, e.g.\
embedded autonomous-vehicle/robot-swarm toolchains
\cite{nouri2025simulation,zhu2025online}, where an external guard
model is an exfiltration risk or infeasible under tactical
compute/connectivity constraints, though we have not evaluated our
probes in such an environment.

Table~\ref{tab:paramcompare} situates our probe's parameter count
against representative external code-vulnerability and content-safety
detectors from the literature (\S\ref{sec:relatedwork}): encoder-only
code models fine-tuned for vulnerability classification (CodeBERT,
GraphCodeBERT) and instruction-tuned LLM guard models (Llama Guard 3,
Granite Guardian, WildGuard). These external sizes are
literature-reported, not benchmarks we ran ourselves
(\S\ref{sec:limitations}), so the comparison is parameter-count only.

\begin{table}[t]
\centering
\footnotesize
\caption{Parameter Count: Probe vs.\ External Detection Models}
\label{tab:paramcompare}
\begin{tabular}{@{}llc@{}}
\toprule
Model & Role & Params \\
\midrule
\textbf{Our Probe} (4 models, Table~\ref{tab:architecture}) & inline, generation-time & 13.41--16.03M \\
\midrule
CodeBERT \cite{feng2020codebert} & fine-tuned classifier & 125M \\
GraphCodeBERT \cite{guo2021graphcodebert} & fine-tuned classifier & 125M \\
Llama Guard 3-8B \cite{metallamaguard3} & external guard model & 8B \\
Granite Guardian 3.x-8B \cite{padhi2024graniteguardian} & external guard model & 8B \\
WildGuard \cite{khatri2026safety} & external guard model & 7B \\
GPT-4-class judge \cite{khatri2026safety} & LLM-as-judge & $\sim$1.8T \\
\bottomrule
\end{tabular}
\end{table}

Our probes are roughly 8--9$\times$ smaller than CodeBERT/GraphCodeBERT,
437--597$\times$ smaller than the 7--8B guard models we compare
against, and five orders of magnitude smaller than a GPT-4-scale
judge. Unlike CodeBERT/GraphCodeBERT, which are fine-tuned end-to-end,
our MLP head reads \emph{frozen} activations off a model the pipeline
needs anyway for generation -- that distinction, not just the
parameter-count gap, drives the latency/infrastructure savings above.
Table~\ref{tab:results} shows this recoverability holds cleanly on
Devign and less so on the three harder benchmarks; the parameter-count
gap, not a measured one, is the evidence we currently have for the
latency argument (\S\ref{sec:limitations}).

\subsection{Relation to Prior Probing Work}
\S\ref{sec:relatedwork} details how our design differs from
\cite{ibanezlissen2025lpass,sternfeld2026minimal}. Neither reports a
directly comparable F1: LPASS targets pruning, and Sternfeld et al.\
report a completion-flip rate under prompt perturbation, not F1 on a
held-out corpus. Our Devign result (68.8\% F1, matching
fine-tuned-classifier SOTA) is the first quantitative evidence we are
aware of that a probe on \emph{unmodified} activations recovers a
comparable signal.

\subsection{Threats to Generalization}
\label{sec:relatedwork-malware}
Probing-based detectors are known to overfit to in-distribution
artifacts: Wang et al. \cite{wang2025falsesense} show malicious-input
probes that look accurate in-distribution fail out-of-distribution,
and Fomin \cite{fomin2026benchmarkslie} shows standard splits can
overstate generalization relative to a true distribution-shift
evaluation. One narrow, encouraging signal: the per-CWE breakdown
(Table~\ref{tab:cwe-accuracy}) agrees across all four models on which
CWE type is hardest -- shared, not per-model idiosyncratic -- though
this is single-dataset agreement, not the cross-dataset test
\cite{wang2025falsesense,fomin2026benchmarkslie} call for. A
leave-one-dataset-out probe and an obfuscation check
\cite{li2025obfuscation} are our highest-priority follow-ups
(\S\ref{sec:limitations}).

\section{Limitations}
\label{sec:limitations}
\begin{itemize}
\item Coverage: 4 models, 3 families (Qwen3.5-9B/Qwen3.6-27B share a
family); not validated outside these families (e.g., closed-weight
models where activations are inaccessible).
\item Final layer, last-prefill-token only, not a layerwise sweep, per
\cite{ibanezlissen2025lpass}. Binary label only; the per-CWE breakdown
(\S\ref{sec:results}) is a post-hoc diagnostic, not a per-CWE risk
score, and runs on Big-Vul only.
\item Dataset caveats (Table~\ref{tab:datasets}): Big-Vul's
spurious-correlation risk, Draper VDISC's imperfect static-analyzer
ground truth, and PrimeVul's severe imbalance ($\sim$3\% vulnerable)
are the main ones. All evaluation is on existing corpus functions, not
a model-generated completion slice (\S\ref{sec:methodology}), and
in-distribution results may not bound worst-case or obfuscated
\cite{li2025obfuscation} performance \cite{wang2025falsesense}.
\item No static analyzer, fine-tuned classifier, or LLM-judge baseline
run ourselves -- Tables~\ref{tab:datasets}/\ref{tab:paramcompare} use
literature-reported numbers, not a controlled head-to-head -- and no
measured wall-clock probe latency (\S\ref{sec:resource-savings}).
\item No architecture ablation: the 6-layer MLP is inherited from
\cite{khatri2026safety}, not independently justified; no linear
(logistic-regression) probe tested (\S\ref{sec:trainingprocedure}),
nor a non-LLM baseline (e.g.\ TF-IDF/AST) to test whether dataset
quality, not model choice, really caps performance.
\item No adversarial/evasion analysis, despite motivating this work by
noting post-hoc detectors are evadable \cite{li2025obfuscation}.
\item Each dataset uses its own native split (\S\ref{sec:setup}), not
a shared, re-stratified one; several Table~\ref{tab:results} cells use
an untuned threshold ($^\dagger$), so within-column rankings should be
read with that caveat.
\end{itemize}

\section{Conclusion}
We investigated whether an LLM's activations over a piece of code
already encode a signal predictive of that code's vulnerability status.
Extending the activation-probing methodology of \cite{khatri2026safety}
from prompt-harm to code vulnerability, we trained MLP probes on four
models (Granite-4.1-8B, Qwen3.5-9B, Qwen3.6-27B, Gemma-4-12B) and four
datasets, achieving 41.7\% average F1 with 13.41--16.03M-parameter
probes -- 0.06--0.17\% of base-model size. Probes match published SOTA
on Devign (up to 68.8\% vs.\ 67.9\% F1) and are remarkably consistent
across models on the harder, more imbalanced benchmarks; we read this
as consistent with dataset quality being the binding constraint there,
though confirming it needs a non-LLM baseline we have not yet run
(\S\ref{sec:limitations}). This is early, preliminary evidence that
vulnerability signal is recoverable by a lightweight probe from a
frozen coding LLM's own activations over existing code -- narrower
than, but a precondition for, a probe recognizing its \emph{own}
forthcoming output as vulnerable. A linear-probe ablation, a non-LLM
baseline, a model-generated-completion pilot, an adversarial-robustness
check, and a cross-model/cross-dataset matrix
(\S\ref{sec:limitations}) are the controls this claim still needs.

\footnotesize
\bibliographystyle{IEEEtran}
\bibliography{references}

\end{document}